\documentclass[superscriptaddress,twocolumn]{revtex4-1}

\usepackage{graphicx}
\usepackage{dcolumn}
\usepackage{bm}

\usepackage[latin1]{inputenc}
\usepackage{amsmath}
 \usepackage{ulem}

\begin{document}

\title{An ideal experiment to determine the `past of a particle' in the nested Mach-Zehnder Interferometer}
\author{Fu Li}%
\affiliation{%
Beijing Computational Science Research Center, Beijing 100084, P.R. China}
\author{F.A. Hashmi}
\affiliation{%
Beijing Computational Science Research Center, Beijing 100084, P.R. China}
 \affiliation{Department of Physics, COMSATS Institute of Information Technology,  Islamabad, Pakistan.}
 \author{Jun-Xiang Zhang}%
\affiliation{%
The State Key Laboratory of Quantum Optics and Quantum Optics Devices, Institute of Opto-Electronics, Shanxi University, Taiyuan 030006, China}

\author{Shi-Yao Zhu}%
\affiliation{%
Beijing Computational Science Research Center, Beijing 100084, P.R. China}
\affiliation{%
The State Key Laboratory of Quantum Optics and Quantum Optics Devices, Institute of Opto-Electronics, Shanxi University, Taiyuan 030006, China}
\affiliation{
Hefei National Laboratory for Physical Sciences at Microscale and Department of Modern Physics,
University of Science and Technology of China, Hefei, Anhui 230026, China}
 \date{\today}
\keywords{Two State Vector Formalism, weak measurements, interference effects}
\pacs{03.65.Ta,03.65.Ca,42.25.Hz}
\newcommand{\ket}[1]{\left|#1\right\rangle}
\newcommand{\bra}[1]{\left\langle#1\right|}
\newcommand{\exptV}[1]{\left\langle#1\right\rangle}
\newcommand{\abs}[1]{\left|#1\right|}
\newcommand{\po}[2]{\left\langle#1|#2\right\rangle}

\newcommand{\op}[2]{\ket{#1}\bra{#2}}

\newcommand{\opR}[1]{\hat{#1}}
\newcommand{\paren}[1]{\left(#1\right)}
\newcommand{\saren}[1]{\left[#1\right]}
\begin{abstract}
An ideal experiment is designed to determine the past of a particle in the nested Mach-Zehnder interferometer (MZI) by using standard quantum mechanics with quantum non-demolition measurements. We find that when the photon reaches the detector, it only follows one arm of the outer interferometer and leaves no trace in the inner MZI; while when it goes through the inner MZI, it cannot reach the detector. Our result obtained from the standard quantum mechanics is contradict to the statement based on two state vector formulism, ``the photon did not enter the (inner) interferometer, the photon never left the interferometer, but it was there''. Therefore, the statement and also the overlap claim are incorrect.
\end{abstract}
\maketitle
One obstacle to describe the past of a quantum particle is the inability to verify any prediction about the past of the particle, as any measurement to observe the particle's path information would cause the wavefunction to collapse at the time of the measurement. Early discussions on path information in quantum mechanics relied on the concept of the duality\cite{Scully91,Herzog95,Jacques07}, which tells us that the price to pay for acquiring the path information is a loss of interference. This old problem of describing the past of a quantum particle has recently resurfaced \cite{Vaidman07, Danan13}, due to the development of weak measurements \cite{Tamir13,Story91,Oreshkov05, Kocsis11, Lundeen11, Steinberg10, Dixon09, Hosten08, Pryde05, Resch04} that do not cause the complete collapse of the wavefunction, and due to counterfactual predications on the particle path, such as interaction-free measurement\cite{Hosten06, Pau95} and counterfactual communication\cite{Salih13}, which may find application in the technology in near future.

In the discussion on the path, an approach, called the two state vector formalism (TSVF) to study the quantum systems between two strong measurements, was proposed \cite{Aharonov90, Aharonov91}. The TSVF makes use of the forward and backward evolving wavefunctions, starting at the time of the pre-selection and at the time of the postselection, respectively. The authors who put forward the TSVF\cite{Vaidman14,Vaidman13,Vaidman07} claimed in \cite{Vaidman14} that ``the particle was in the overlap of the forward and backward evolving quantum states''. Based on this claim, they stated for the nested Mach-Zanhder interferometer (MZI) that,``we can state the following: the photon did not enter the (inner) interferometer, the photon never left the interferometer, but it was there'' \cite{Vaidman07}. This statement raised serious controversies and led to considerable debate \cite{Li13,VaidmanReply13}. Recently, an experiment \cite{Danan13} was reported where the authors said,``The experimental results have a simple explanation in the framework of the two-state vector formalism (TSVF) of quantum theory'' which means that the experiment supports the claim and statement. However, their experiment itself is controversial \cite{PhysRevA033825,HatimSalih,Bengt}, and Ref.\cite{PhysRevA033825} clearly showed that the experiment did not prove that the statement is correct. The main point of contention is that the weak measurement will destroy the destructive interference at the dark port and cause a leakage through the dark port. That leakage contributes the photon's trace in the inner interferometer revealed at detector D. As a consequence the weak measurement cannot resolve the controversy. In this letter, an experiment, that uses quantum non-demolition (QND) measurements in the nested MZI to reveal whether the photon is presented in the inner MZI if the detector D clicks, is proposed. The novelty of our scheme is that we extract the path information without disturbing the quantum interference (the dark port remaining dark). Our results based on the standard quantum mechanics but not the TSVF, show that the statement in\cite{Vaidman07} is incorrect. As the direct conclusion from the overlap claim based on the TSVF is incorrect, the TSVF itself should be re-examed.

\begin{figure}[htbp] \begin{center}
 \includegraphics[scale=0.33]{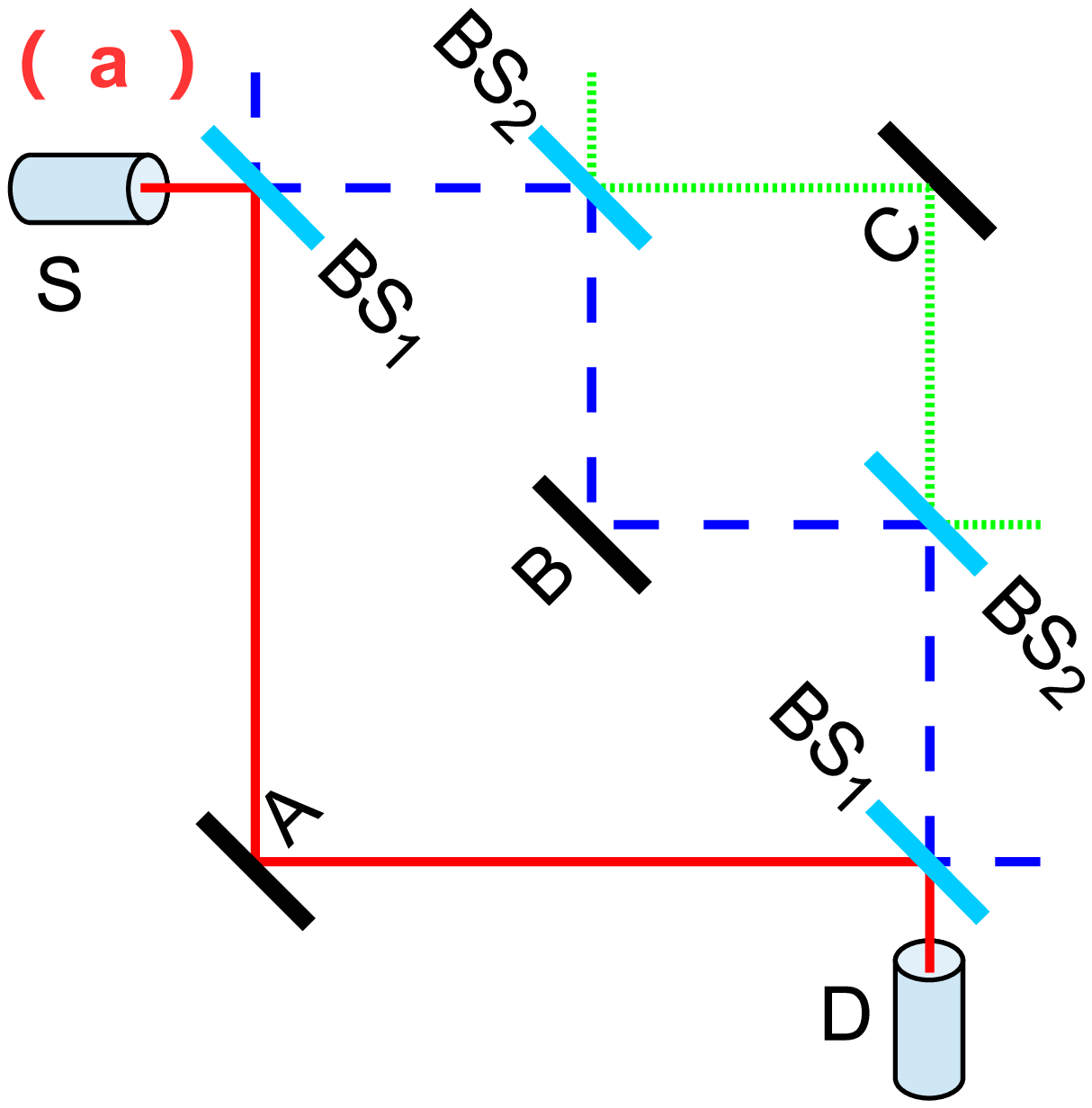}
 \includegraphics[scale=0.33]{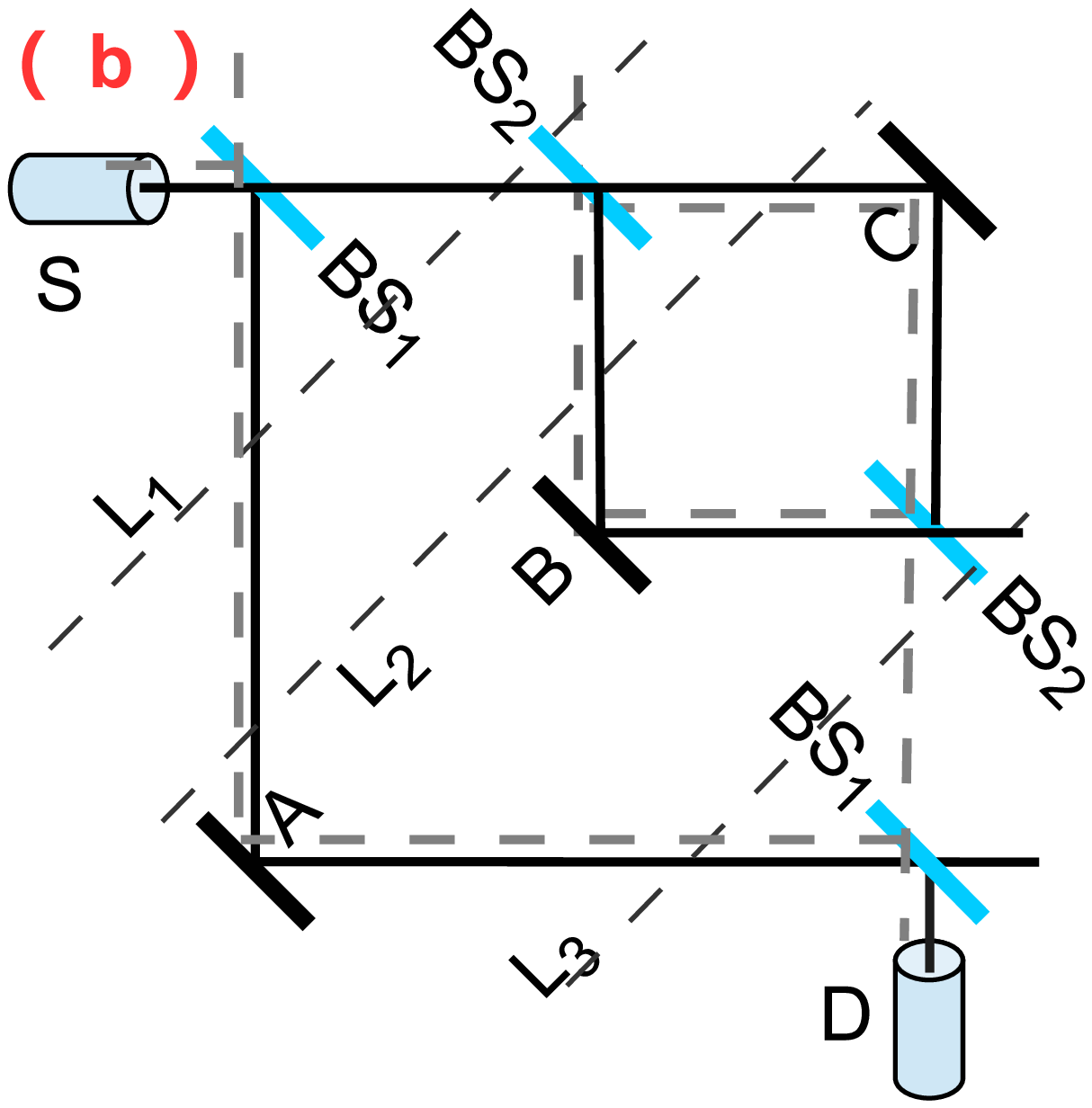} \end{center}
 \caption{The nested Mach-Zehnder interferometer. We use column vectors $( \begin{smallmatrix} n_1 &n_2&n_3 \end{smallmatrix})^\dagger$ to describe the state of the system, where $n_1$, $n_2$, and $n_3$ are the
number of photons along the modes$(\begin{smallmatrix}1 &0&0 \end{smallmatrix})^\dagger$(red),$(\begin{smallmatrix}0&1&0 \end{smallmatrix})^\dagger$(blue), and $(\begin{smallmatrix}0 &0&1 \end{smallmatrix})^\dagger$(green), respectively. The lines L1, L2, and L3 in Fig.1b represent different stages during the evolution. } \label{fig1} \end{figure}

The nested interferometer consists of an outer bigger interferometer with two beam splitters $\mathrm{BS}_1$, and an inner interferometer, comprising of the beam splitters $\mathrm{BS}_2$ along one arm of the outer interferometer. The beam splitters $\mathrm{BS}_1$ have reflectivity $r$ and transmissivity $t$, whereas the beam splitters $\mathrm{BS}_2$ are $50:50$. This arrangement  makes the output port of the inner interferometer towards the detector D a dark port, and the wavefunction from inside the inner interferometer can not reach the detector. In this setup,
we have three quantum paths which are indicated by red solid, blue dashed and green dotted lines in Fig. 1(a). We use column vectors $( \begin{smallmatrix} n_1 &n_2&n_3 \end{smallmatrix})^\dagger$ to describe the state of the system, where $n_1$, $n_2$, and $n_3$ are the
number of photons along the modes$(\begin{smallmatrix}1 &0&0 \end{smallmatrix})^\dagger$(red),$(\begin{smallmatrix}0&1&0 \end{smallmatrix})^\dagger$(blue), and $(\begin{smallmatrix}0 &0&1 \end{smallmatrix})^\dagger$(green), respectively.  We also define the corresponding photon creation and annihilation operators $\opR{a}_i^\dagger$ and $\opR{a}_i$ ($i=1,2,3$), respectively. The lines L1, L2, and L3 in Fig.1b represent different stages during the evolution. Stage L1 is between the first $\mathrm{BS}_1$ and the first $\mathrm{BS}_2$, L2 between the  two $\mathrm{BS}_2$,  and L3 is between the second $\mathrm{BS}_2$ and the second $\mathrm{BS}_1$.

	If a single photon coming from the source $\mathrm{S}$ has been detected by D, the pre-selection is $\ket{\psi}= (\begin{smallmatrix}1&0&0\end{smallmatrix})^\dagger$, and the post-selected state is $\bra{\phi}=(\begin{smallmatrix}1&0&0\end{smallmatrix})$, in the TSVF. The pre-selected wavefunction evolves forward in time through different beam splitters following the black solid line, see Fig.1b.  At stage L1, the wavefunction is $\ket{\psi_{L1}}= (\begin{smallmatrix}-ir&t&0\end{smallmatrix})^\dagger$ . At stage L2 the photon wavepacket is present both along the arm A of the outer interferometer and inside the inner interferometer, $\ket{\psi_{L2}}=(\begin{smallmatrix}-ir&-it/\sqrt{2}&t/\sqrt{2}\end{smallmatrix})^\dagger$ . Due to the dark port of the inner interferometer along the second mode $(\begin{smallmatrix}0&1&0\end{smallmatrix})^\dagger$  (see blue dashed line in Fig.1a), the wavefunction at stage L3 is  $\ket{\psi_{L3}}= (\begin{smallmatrix}-ir&0&it\end{smallmatrix})^\dagger$. This shows that the particle which was inside the inner interferometer leaves the system along the mode $(\begin{smallmatrix}0&0&1\end{smallmatrix})^\dagger$ and cannot contribute to the post-selection at D.

The backward evolving wavefunction (the post-selected state) created at the detector after the successful photon detection evolves backward following the grey dashed line. This backward evolution can be through arm A and the inner interferometer. However, the portion passing through the inner interferometer will leave the system at stage L1 and cannot reach the source (because of the dark port), see grey dashed line in Fig. (1b). Based on the TSVF, the photon in its past should be present at the places where the two wave-functions overlap, which includes the arm A of the outer interferometer and the inner interferometer, but not the paths leading to and coming out of the inner interferometer \cite{Vaidman14,Vaidman13,Vaidman07}.

\begin{figure}[htbp] \begin{center} \includegraphics[scale=0.6]{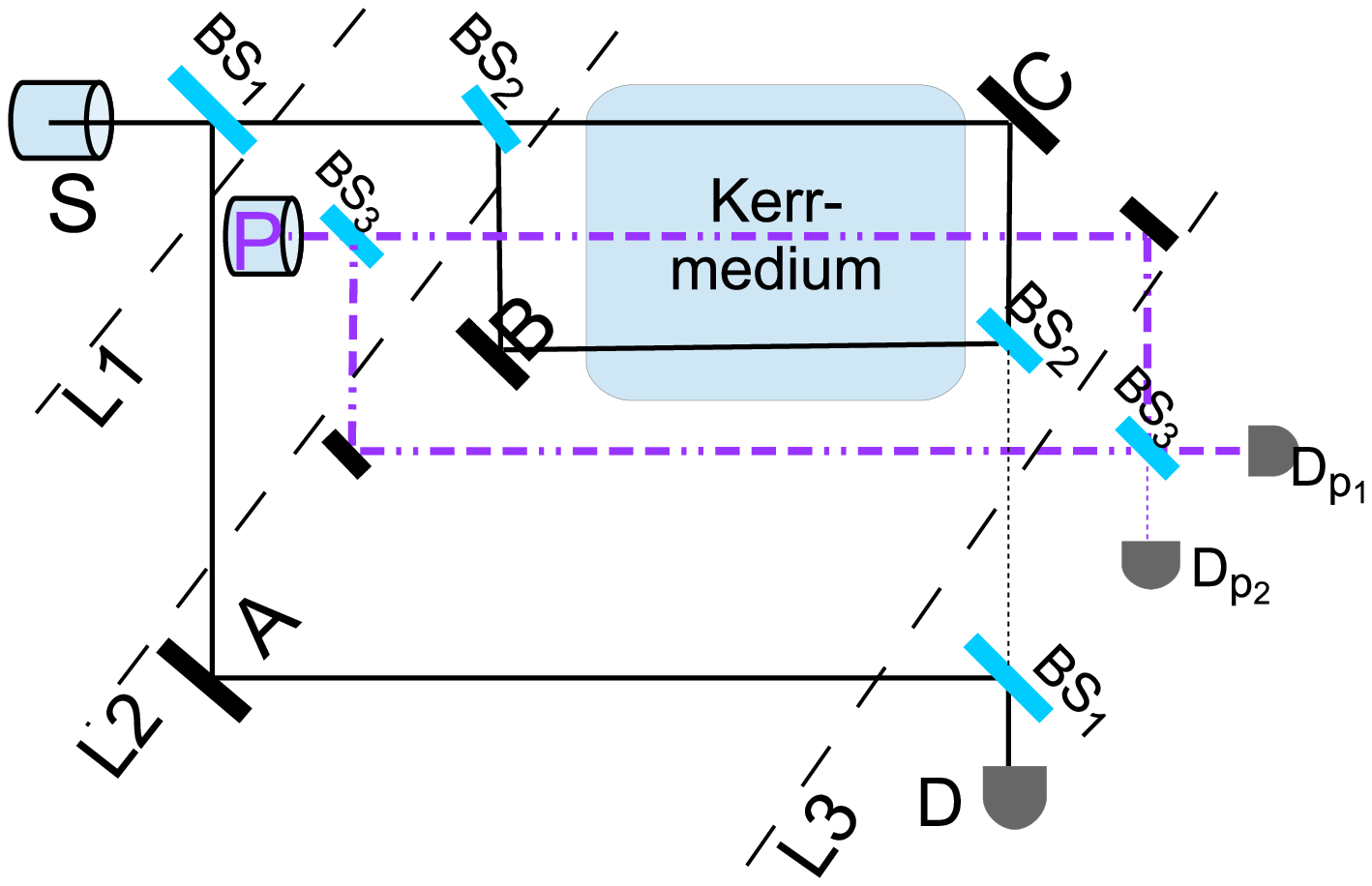} \end{center} \caption{Non-demolition(QND) measurement that can reveal the presence of the photon inside inner interferometer without disturbing its interference, two paths of the inner interferometer are symmetric with respect to the middle beam in the Kerr-medium.} \label{fig2} \end{figure}

	In order to test the statement that the photon detected by D was present in the inner interferometer, we design an experiment with QND measurement\cite{Barrett05,Imoto85,Munro05,Feizpour11}. The novelty of the setup is that we can probe the presence of the photon inside the inner interferometer without disturbing the destructive interference on the dark port. A third interferometer and a coherent field is added as a
probe to reveal the photon trace in the inner interferometer. The coherent state $\ket {\sqrt{2}\alpha}$ is split by a $50:50$ beam splitter ($\mathrm{BS}_3$) into two coherent fields, $\ket{\alpha}_1$ and $\ket{i\alpha}_2$, which enter the two arms of the third interferometer, respectively. A Kerr-media is placed along the two paths of the inner interferometer and one arm of the third interferometer. This arm of the third interferometer is placed in the middle of two paths of the inner interferometer, see  Fig.~\ref{fig2}. The output of the third interferometer ($\mathrm{D}_{\mathrm p1}$ and $\mathrm{D}_{\mathrm p2}$) will give us a fringe pattern due to the interference between two coherent fields. The fringe pattern detected by $\mathrm{D}_{\mathrm p1}$ (or $\mathrm{D}_{\mathrm p2}$) in the case of the photon passing through the inner interferometer (due to the interaction with the coherent fields) is different from that in the case of the photon not passing through the inner interferometer. Thus from the fringe pattern, we can determine whether the photon passes through the inner interferometer

The arm of the third interferometer that passes through the Kerr-medium carries the field ($\ket{\alpha}_1$).
Inside the Kerr-medium, the interaction between the coherent beam and the photon inside the inner interferometer can be represented by the Hamiltonian\cite{Zubairy02,Imoto85},
\begin{align}
 H = \varepsilon \sum\limits_{i = 2,3} {\opR{a}_i^\dag {\opR{a}_i}\opR{a}_{p1}^\dag {\opR{a}_{p1}}}  + \eta \opR{a}_2^\dag {\opR{a}_2}\opR{a}_3^\dag {\opR{a}_3} \label{eqHamiltonian2}
 \end{align}
	where $\opR{a}_{p1}^\dagger$ and $\opR{a}_{p1}$ are the creation and annihilation operators for the photons of  $\ket{\alpha}_1$, and $\eta$ is the interaction strength between the two paths of the inner interferometer in the Kerr-medium. As the two paths of the inner interferometer are symmetric with respect to the middle coherent state $\ket{\alpha}_1$,the measurement interaction strength between the two paths of the inner interferometer with the coherent field in the Kerr-medium are same, noted with $\epsilon$.

 With the pre-selection $\ket{\psi_i} =(\begin{smallmatrix}1&0&0\end{smallmatrix})^\dagger$, the joint state of the system and the coherent fields at the stage L2 (before the interaction) is given as
\begin{align}\label{eqL2} \ket{\psi_{L2}}= \begin{pmatrix}ir \\0 \\ 0\end{pmatrix} \ket{\alpha}_1\ket{i\alpha}_2 +\begin{pmatrix}0 \\it/\sqrt{2} \\t/\sqrt{2} \end{pmatrix} \ket{\alpha}_1\ket{i\alpha}_2 \end{align}
After the interaction with the  Kerr-media, the joint state is
\begin{align}\label{eqL22} \ket{{\psi^{\prime}}_{L2}}= \begin{pmatrix}ir \\0 \\ 0\end{pmatrix} \ket{\alpha}_1\ket{i\alpha}_2 +e^{ - i\eta {\tau_0}}\begin{pmatrix}0 \\it/\sqrt{2} \\t/\sqrt{2}\end{pmatrix} \ket{\alpha e^{-i\epsilon \tau_0}}_1\ket{i\alpha}_2
\end{align}
 where we have set the passing time of the two arms through the Kerr-medium are the same, $\tau_1=\tau_2=\tau_0$ . The joint state of the system and the coherent fields at the stage L3 is given as
\begin{align}\label{eqL3}
\left| {{\psi _{L3}}} \right\rangle  = \left( {\begin{array}{*{20}{c}}
  {ir} \\
  0 \\
  0
\end{array}} \right){\left| \alpha  \right\rangle _1}{\left| {i\alpha } \right\rangle _2} + {e^{ - i\eta {\tau_0}}}\left( {\begin{array}{*{20}{c}}
  0 \\
  0 \\
  {it}
\end{array}} \right){\left| {\alpha {e^{ - i\varepsilon {\tau_0}}}} \right\rangle _1}{\left| {i\alpha } \right\rangle _2}
\end{align}
Note that the second mode $(\begin{smallmatrix}0&1&0\end{smallmatrix})^\dagger$ is still empty. This is the consequence of not exploring the `which-path' information in the inner interferometer. It is essential that $\tau_1=\tau_2=\tau_0$, so that the dark port of the inner MZI kept still dark after the Kerr medium being added, which can be achieved by paralleling the left and right edges of the Kerr medium (see Fig.2). Paralleling the two edges can be realized in experiment using current technology.

 After the second $\mathrm{BS}_3$ and before the second $\mathrm{BS}_1$, the joint state is
\begin{align}
 \ket{{\psi^\prime}_{L3}}= \begin{pmatrix}ir \\ 0 \\ 0\end{pmatrix} \ket{i\sqrt{2}\alpha}_1\ket{vacuum}_2 \qquad\qquad\qquad\nonumber\\
  +e^{-i\eta \tau_0}\begin{pmatrix}0 \\ 0 \\it \end{pmatrix} \ket{i\alpha\frac{ 1+e^{-i\epsilon \tau_0}}{\sqrt{2}}}_1\ket{\alpha \frac{ 1-e^{-i\epsilon \tau_0}}{\sqrt{2}}}_2 \label{eqL3}
\end{align}
Although we do not know the `which-path' information of the photon in the inner interferometer, we can still determine whether the photon passed through the inner interferometer. Only the first term in Eq. (5) contain the system photon mode that can reach the detector D. However, it is clear from this term that the photon wavefunction reaching the detector D has not interacted with the probe coherent field and has left no trace on the fringes in the detectors $\mathrm{D}_{\mathrm p1}$ and $\mathrm{D}_{\mathrm p2}$. The second term in Eq. (\ref{eqL3}) describes the portion of the wavefunction that has interacted with the photon inside the inner interferometer and has left a trace (a shift) on the fringes, but this portion of the photon wavefunction leaves the system along the mode  $( \begin{smallmatrix} 0&0&1 \end{smallmatrix})^\dagger$  at the stage L3, and never reaches the detector D. It clearly proves that the photon that has been detected at the detector D was following only the arm A of the outer interferometer. It was not inside the inner interferometer and has not left any trace inside the inner interferometer. This straight forward quantum mechanical reasoning is in clear contradiction with the prediction of TSVF that associates the past of the photon with the overlap of the forward and backward evolving waves and thus with the inner interferometer. In the standard quantum mechanics, forward evolving wavefunction is enough to describe the whole evolution of the system.

Let us tentatively use the backward evolving wavefunction of TSVF. Suppose that the back evolution state including the coherent state is $\bra{\phi_{f}}=\begin{pmatrix}1&0&0\end{pmatrix}\bra{-i\sqrt{2}\alpha}_1\bra{vacuum}_2$ . We can derive from the standard quantum mechanics the back evolution states at different stages,

\begin{align}
\bra{\phi_{L3}}=\begin{pmatrix}ir&t&0\end{pmatrix}\bra{\alpha}_1\bra{-i\alpha}_2
 \end{align}

 \begin{align}
\bra{\phi_{L2}}=&\begin{pmatrix}ir&0&0\end{pmatrix}\bra{\alpha}_1\bra{-i\alpha}_2 \\ \nonumber
                            &+e^{-i\eta \tau_0}\begin{pmatrix}0&it/\sqrt{2}&t/\sqrt{2}\end{pmatrix}\bra{\alpha e^{-i\varepsilon \tau_0}}_1\bra{-i\alpha}_2
 \end{align}

 \begin{align} \label{eqbkL1}
\bra{\phi_{L1}}=&\begin{pmatrix}ir&0&0\end{pmatrix}\bra{ \sqrt{2}\alpha}_1\bra{ vacuum}_2\\ \nonumber
                            &+e^{-i\eta \tau_0}\begin{pmatrix}0&0&it\end{pmatrix}\bra{\frac{1+e^{-i\varepsilon \tau_0}}{\sqrt{2}}\alpha}_1\bra{-i\frac{1-e^{-i\varepsilon \tau_0}}{\sqrt{2}}\alpha}_2
 \end{align}

From the first term in Eq. (\ref{eqbkL1}), we can clearly see that if the single photon evolves back to the pre-selection, so does the coherent state, which means the single photon leaves no trace in the inner interferometer (and on the measurement device). The second term tells us that
part of the coherent state does not evolve to the pre-selection $\bra{ \sqrt{2}\alpha}_1$, and the measurement device gives us the information about the system. The second term is the result of the corresponding single photon that leaves a trace in the inner interferometer and then goes away, and will not reach the source S. Hence, the backward evolving wavefucntion tells the same story as the forward evolving wavefunction. A particle going back from the detector D to the source S can not leave a trace inside the inner interferometer.

In above, we propose an ideal experiment for the nested MZI system. In the experiment, the quantum non-demolition (QND) measurements are used to reveal the past of the quantum particle without disturbing the interference of the system (keeping the dark port still dark), which are different from the weak measurement in \cite{Vaidman07, Danan13} that disturbs the interference (leading to a leakage to the dark port). Our derivation, based on the standard quantum, shows that the photon was only following path A and leaves no trace in the inner interferometer when the detector D (post-selection) has a click. On the opposite, when the photon passes through the inner interferometer, the detector D has no click. This conclusion is contradicted with the statement and the overlap claim from the TSVF. Please note that the overlap claim of the TSVF is not derived from or a result of the standard quantum mechanics. Therefore, the contradictory between the overlap claim of the TSVF and the standard quantum mechanics means the overlap claim is incorrect. Our conclusion can also remove the doubts or settle the argument on the counterfactual communication\cite{Salih208902,Vaidman208901,Vaidman07,Salih13,Pau95}. In our proposed experiment, the phase shift is proportional to the nonlinear coefficient, $\chi(3)$. Large amount of research work is focused on searching materials of large $\chi(3)$ \cite{Zubairy02,Ottaviani03,Gain07,Thibault12}, and controlling the shift from zero to $\pi$ \cite {LiRB13}. Measurable shift with a weak probe field \cite{Kang03,Lo10,Lo11}, even at the singe photon level \cite{Thibault12}, was experimentally observed. We are hopeful that in the near future, the proposed experiment would be realized with the development of experiment technology.
 \begin{acknowledgements}
This work was supported by the National Basic Research Program
of China (Grants No.~2011CB922203 and No.~2012CB921603)
and the National Natural Science Foundation of China
(Grants No.~1174026 and No.~U1330203).
\end{acknowledgements}
\bibliographystyle{apsrev}
\bibliography{paper}

\end{document}